\documentclass[prl,twocolumn,showpacs,preprintnumbers,amsmath,amssymb,superscriptaddress, bibnotes]{revtex4}

\usepackage{color}
\usepackage{graphicx}
\usepackage{dcolumn}
\usepackage{bm}

\begin{document}

\title{Ferromagnetic Quantum Critical Point in Heavy-Fermion Iron Oxypnictide Ce(Ru$_{1-x}$Fe$_x$)PO}

\author{S.~Kitagawa}
\email{shunsaku@scphys.kyoto-u.ac.jp}
\author{K.~Ishida}
\affiliation{Department of Physics, Graduate School of Science, Kyoto University, Kyoto 606-8502, Japan}

\author{T.~Nakamura}
\author{M.~Matoba}
\author{Y.~Kamihara}
\affiliation{Departments of Applied Physics and Physico-Informatics, Keio University, Kanagawa, 223-8522, Japan}

\date{\today}

\begin{abstract}
We have performed $^{31}$P-NMR measurements on  Ce(Ru$_{1-x}$Fe$_{x}$)PO in order to investigate ferromagnetic (FM) quantum criticality, since heavy-fermion (HF) ferromagnet CeRuPO with a two-dimensional structure turns to a HF paramagnet by an isovalent Fe-substitution for Ru.       
We found that Ce(Ru$_{0.15}$Fe$_{0.85}$)PO shows critical fluctuations down to $\sim 0.3$ K, as well as the continuous suppression of Curie temperature and the ordered moments by the Fe-substitution. 
These experimental results suggest the presence of a FM quantum critical point (QCP) at $x \sim 0.86$, which is a rare example among itinerant ferromagnets.
In addition, we point out that the critical behaviors in Ce(Ru$_{0.15}$Fe$_{0.85}$)PO share the similarity with those in YbRh$_2$Si$_2$, where the local criticality of $f$ electrons has been discussed.
We reveal that Ce(Ru$_{1-x}$Fe$_{x}$)PO is a new system to study FM quantum criticality in HF compounds. 
\end{abstract}


\abovecaptionskip=-5pt
\belowcaptionskip=-10pt

\maketitle

It has been believed that a ferromagnetic (FM)- paramagnetic (PM) transition at zero magnetic field is a text-book example of a 2nd-order phase transition at finite temperatures\cite{J.A.Hertz_PRB_1976}.
However, there have been a lot of reports that the FM-PM transition changes from 2nd-order to 1st-order at a tricritical point (TCP) in metallic ferromagnets, when the FM transition is suppressed to 0 K by non-thermal control parameters\cite{M.Uhlarz_PRL_2004,C.Pfleiderer_PRB_1997,F.Levy_NaturePhys_2007,H.Kotegawa_JPSJ_2011,K.Adachi_JPSJ_1979}.
In addition, the quantum FM transition was predicted to be generically of 1st-order from the theoretical viewpoints\cite{H.Yamada_PRB_1993,D.Belitz_PRL_2005}. 
Nature of the quantum phase transition from the FM to PM state in metallic ferromagnets has been revisited recently.

A quantum critical point (QCP), defined as a zero-temperature 2nd-order phase transition driven by non-thermal physical parameters, has attracted much attention, since fascinating phenomena, e.g., unconventional superconductivity\cite{G.Knebel_PRB_2006,S.Kasahara_PRB_2010} and nematic order\cite{R.A.Borzi_Science_2007} have been discovered near a QCP. 
For studying the nature of a QCP, heavy-fermion (HF) systems, in which $f$ electrons behave as itinerant electrons with heavy effective-mass, are one of the best systems for this purpose, since a magnetic transition can be tuned by various physical parameters such as chemical substitution\cite{A.Schroder_Nature_2000,P.G.Pagliuso_PRB_2001}, pressure\cite{M.Ohashi_PRB_2003,E.Hassinger_JPSJ_2008} or magnetic field\cite{P.Gegenwart_PRL_2002}. 
Up to now, there have been considerable efforts for understanding the nature of a QCP on various antiferromagnetic (AFM) HF compounds, but very few on FM HF compounds, particularly on Ce-based FM HF compounds. 

The iron oxypnictide CeFe(Ru)PO is a related material of the iron-based superconductor LaFePO\cite{Y.Kamihara_JACS_2006}.
They possess the same two-dimensional layered structure, stacking the Ce(La)O and Fe(Ru)P layers alternatively. 
The CeO layer contributes to the large magnetic response, and the Fe(Ru)P layer is conductive in CeFe(Ru)PO. 
CeRuPO is a FM HF compound with Curie temperature $T_{\rm C}$ = 15 K and coherent temperature $T_{\rm K}$ $\simeq$ 10 K \cite{C.Krellner_PRB_2007}. 
On the other hand, counterpart CeFePO is a HF compound with a PM ground state down to 80 mK and possesses a large Sommerfeld coefficient $\gamma$ = 700 mJ/mol$\cdot$K$^2$ at low temperatures\cite{E.Buning_PRL_2008}. 
Quite recently, we reported that CeFePO shows a metamagnetic (MM) behavior related to the Kondo effect when magnetic fields are applied perpendicular to the $c$ axis, and exhibits a non Fermi liquid behavior at the MM field \cite{S.Kitagawa_PRL_2011}.
Therefore, we expect that the ground state can be tuned continuously from FM to PM by substituting isovalent Fe for Ru. 
Actually, continuous change from FM to PM state revealed by magnetization and specific heat measurements down to 2~K is reported in CeFe(As$_{1-x}$P$_{x}$)O \cite{Y.Luo_PRB_2010}.
In particular, we anticipate that Ce(Ru$_{1-x}$Fe$_{x}$)PO is an ideal system for investigating physical properties of FM quantum criticality from a microscopic point of view by $^{31}$P-NMR measurements, since the substitution is done at the Fe(Ru)P layer and thus introduces small randomness effect at the magnetic CeO layer as well as at the P site detected with NMR. Here we report evolution of magnetic properties in Ce(Ru$_{1-x}$Fe$_{x}$)PO revealed by $^{31}$P-NMR measurements and show the phase diagram in this system. 
We stress that an NMR measurement is one of the best experimental technique for distinguishing between 1st and 2nd order transition as mentioned in our previous paper~\cite{K.Karube_PRB_2012}.

\begin{figure}[tb]
\vspace*{-10pt}
\begin{center}
\includegraphics[width=9cm,clip]{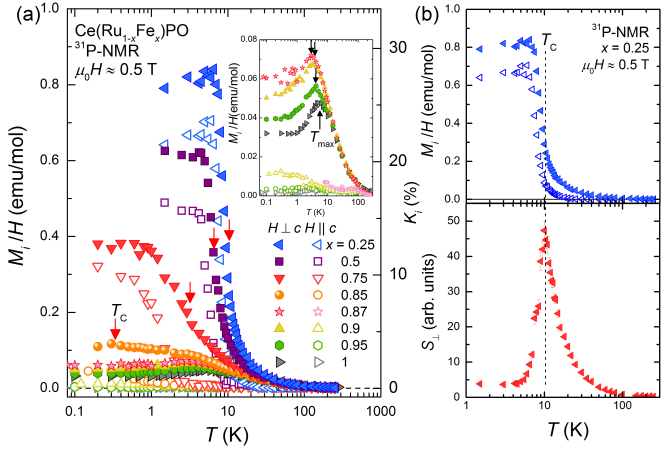}
\end{center}
\caption{(Color online)(a) $T$ dependence of $M_i/H$ evaluated from the Knight shift determined at the peaks of $H \perp c$ and $H \parallel c$ spectra obtained at $\simeq$ 0.5~T for Ce(Ru$_{1-x}$Fe$_{x}$)PO ($x$~=~0.25, 0.5, 0.75, 0.85, 0.87, 0.9, 0.95, and 1). The Knight shift $K_{\perp}$ and $K_c$ are also represented in the right axis. $M_i/H$ is estimated from the equation $M_i/H$ = $(K_i - K_{\rm orb})/A_{\rm hf,iso}$ where isotropic hyperfine coupling constant $^{31}A_{\rm hf,iso}$ = 0.2 T/$\mu_B$ and $T$-independent orbital part of Knight shift $K_{\rm orb}$ = 0.09 \% derived at $x = 1$ were used. Below $x = 0.85$, $K_i$ increases below $T_{\rm C}$ due to the appearance of the internal field at the P site. The arrows indicate $T_{\rm C}$. 
(Inset) $T$ dependence of $M_i/H$ above $x = 0.87$. 
$M_{\perp}/H$  shows a peak at around $T_{\rm max}$ $\sim$ 5~K (indicating the arrows) instead of the FM order, suggestive of the PM HF ground state at low fields. 
(b) $T$ dependence of $M_i/H$ (upper panel) and in-plane dynamical susceptibility $S_{\perp}$ probed by $1/T_1$ (lower panel) at $x = 0.25$. $T_{\rm C}$ (broken line) is unambiguously determined by the peak of $S_{\perp}$.}
\label{Fig.1}
\end{figure}

The polycrystalline Ce(Ru$_{1-x}$Fe$_{x}$)PO ($x$~=~0.25, 0.5, 0.75, 0.85, 0.87, 0.9, 0.95, and 1) was synthesized by solid-state reaction~\cite{Y.Kamihara_JPCS_2008,T.Nakamura_ICM_2012}. 
As reported in the previous paper\cite{S.Kitagawa_PRL_2011}, these possess a two-dimensional anisotropy, so the polycrystalline samples were uniaxially aligned by taking advantage of the anisotropy of the magnetic susceptibility. 
The $c$ axis of the samples are mostly aligned, but the $a$ and $b$ axes are randomly oriented in the $ab$ plane. $^{31}$P-NMR measurements were performed on the aligned  samples. 

NMR spectra were obtained with the field-swept technique, and the Knight shift, which is the measure of the local susceptibility at the nuclear site, was determined from the peak field of each resonance spectrum. Knight shift $K_i$($T$, $H$) ($i = \perp$ and $c$) is defined as,  
\begin{equation}
K_i(T, H) = \left(\frac{H_0 - H_{\rm res}}{H_{\rm res}}\right)_{\omega = \omega_0} \propto \frac{M_i(T,H_{\rm res})}{H_{\rm res}},
\end{equation}
where $H_{\rm res}$ is a magnetic field at a resonance peak, and  $H_0$ and $\omega_0$ are the resonance field and frequency of a bare $^{31}$P nucleus and have the relation of $\omega_0 = \gamma_n H_0$ with the $^{31}$P-nuclear gyromagnetic ratio $\gamma_n$. 
The Knight shift was measured in $H$ parallel ($K_{c}$) and perpendicular ($K_{\perp}$) to the $c$ axis for all samples at $\mu_0H \simeq 0.5$~T.
Reflecting the two-dimensional $XY$-type anisotropy, $K_{c}$ proportional to the out-of-plane component of the susceptibility is small and almost $T$ independent, but $K_{\perp}$  proportional to the in-plane susceptibility shows strong $T$ dependence originating from the Curie-Weiss (CW) behavior of $\chi(T)$ at high temperatures as shown in Fig.~\ref{Fig.1}(a). 
At $x$ = 0.25, $K_c$ suddenly increases below $T_{\rm C}$ = 10~K due to the appearance of the internal field at the P site, but $K_{\perp}$ gradually increases from much higher temperature than $T_{\rm C}$ due to the CW behavior. $T_{\rm C}$ is unambiguously determined by the in-plane low-energy spin fluctuations $S_{\perp}$ as shown in Fig.~\ref{Fig.1}(b) probed by the nuclear spin-lattice relaxation rate ($1/T_1$) in $H \parallel c$. 
As discussed in the previous paper~\cite{S.Kitagawa_PRL_2011}, $1/T_1$ probes spin fluctuations perpendicular to the applied magnetic field, and thus $1/T_1$ in $H \parallel c$ and $H \perp c$ are described as,
\begin{align}
\left(\frac{1}{T_1}\right)_{H\parallel c} \equiv 2S_{\perp}, 
\left(\frac{1}{T_1}\right)_{H\perp c} \equiv S_c + S_{\perp}.
\end{align}
From two equations, we can know low-energy $q$-summed spin fluctuation of in-plane ($S_{\perp}$) and out-of-plane ($S_c$) component, separately.

\begin{figure}[tb]
\vspace*{-10pt}
\begin{center}
\includegraphics[width=8.5cm,clip]{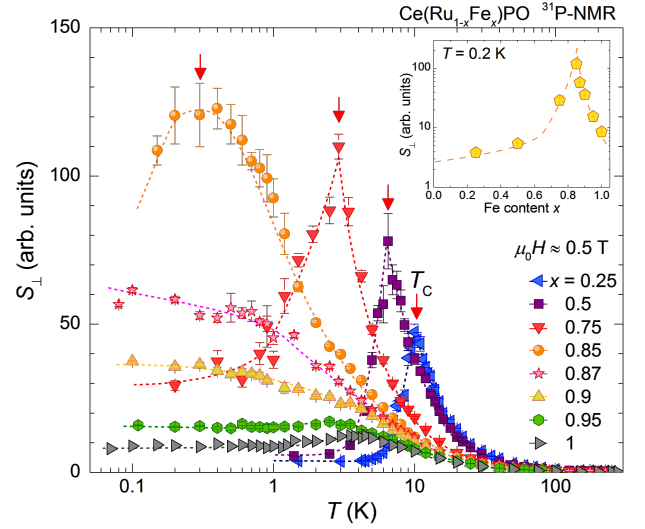}
\end{center}
\caption{(Color online) $T$ dependence of low-energy in-plane spin fluctuations $S_{\perp}$ at $\simeq$ 0.5~T. (See in the text.) Below $x = 0.85$, $T$ dependence of $S_{\perp}$  shows a peak at $T_{\rm C}$ (indicating the arrows). Above $x = 0.9$, $S_{\perp}$ as well as $K_{\perp}$ along both directions becomes almost a constant at low temperatures, indicative of the formation of a HF ground state, although $S_{\perp}$ continue to increase down to 0.08~K at $x = 0.87$. 
(Inset) $x$ dependence of $S_{\perp}$ at 0.2~K. $S_{\perp}$ diverges toward $x \sim 0.86$.
A broken curve is guide to eyes.}
\label{Fig.2}
\end{figure}

By substitution of Fe for Ru, $T_{\rm C}$ and the internal field at the P site, proportional to ordered magnetic moment $\langle \mu_{\rm ord}\rangle$, continuously approach to zero toward $x \sim 0.86$, suggesting that a FM QCP is located at $x \sim 0.86$. 
The $x$ dependence of $S_{\perp}$ also indicates the existence of a FM QCP at $x \sim 0.86$ as shown in Fig.~\ref{Fig.2}.

\begin{figure}[tb]
\vspace*{-10pt}
\begin{center}
\includegraphics[width=8cm,clip]{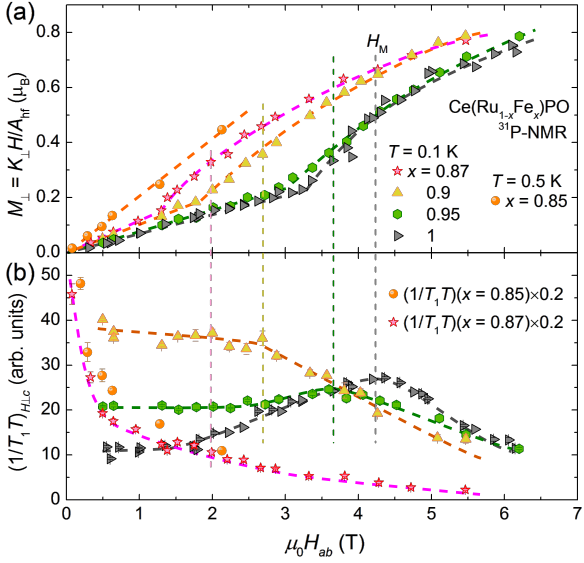}
\end{center}
\caption{(Color online)  (a) $H$ dependence of magnetization $M_{\perp}(H)$ in $H \perp c$ at 0.1~K below $x = 0.87$ and at 0.5~K for $x = 0.85$ just above $T_{\rm C}$ using the relation of  $M_{\perp} (H) = K_{\perp}(H)H/ A_{\rm hf,iso}$. Dashed curves are guide to eyes. 
Above $x = 0.87$, the $M_{\perp}(H)$ suddenly increases with increasing $H$ and deviates from linear relation in the field range of 1 - 5~T, which is a definition of a metamagnetic behavior, although $M_{\perp}(H)$ increase linearly at $x = 0.85$. 
(b) $H$ dependence of $(1/T_1T)_{H \perp c}$ at 0.1~K below $x = 0.87$ and at 0.5~K for $x = 0.85$. 
At $x = 1$, the low-field PM state with HF character is separated from the high-field polarized PM state by a broad peak of $(1/T_1T)_{H \perp c}$ at $H_{\rm M} \simeq 4.3$ T. 
Similar broad peak $(1/T_1T)_{H \perp c}$ was observed at $x = 0.95$, but is changed to a shoulder structure at $x = 0.9$. 
On the other hand, $(1/T_1T)_{H \perp c}$ is monotonically suppressed by magnetic field and no anomaly is observed at $x = 0.85$ and 0.87.
We define $H_{\rm M}$ from the $H$ dependence of $M_{\perp}$ and $(1/T_1T)_{H \perp c}$ as shown by the dotted lines. $H_{\rm M}$ continuously decreases with decreasing $x$. Dashed curves are guide to eyes.}
\label{Fig.3}
\end{figure}

Above $x = 0.9$, spin fluctuation $S_{\perp}$  as well as $K_{\perp}$ becomes almost constant at low temperatures, indicative of the formation of the PM HF state, although $S_{\perp}$ continue to increase down to 0.08~K at $x = 0.87$. 
At $x = 1$, as mentioned above, we reported MM behavior at the MM field $H_{\rm M} \simeq 4.3$~T, and suggested that $H_{\rm M}$ corresponds to an energy breaking the HF state, since the linear relationship holds in various Ce-based HF metamagnets, between $H_{\rm M}$ and the temperature where susceptibility shows a broad maximum $T_{\rm max}$ or the inverse of the effective mass. 
We investigated the $x$ revolution of the MM behavior below $x = 0.85$. 
Figure~\ref{Fig.3}(a) shows the $H$ dependence of magnetization $M_{\perp}$ at 0.1~K below $x = 0.87$ and at 0.5~K for $x = 0.85$ just above $T_{\rm C}$, which are PM state at zero field. 
$M_{\perp}$ is defined as $M_{\perp} = K_{\perp}H/A_{\rm hf,iso}$ with the isotropic hyperfine coupling constant $^{31}A_{\rm hf,iso}$ = 0.2 T/$\mu_{\rm B}$ derived at $x = 1$\cite{S.Kitagawa_PRL_2011}. 
We observed the similar MM behavior at $x = 0.87, 0.9$ and 0.95, and found that $H_{\rm M}$ decreases with decreasing $x$, consistent with the suppression of $T_{\rm max}$ with decreasing $x$. [See in the inset of Fig.~\ref{Fig.1}(a).] 
The MM behavior was also confirmed from the spin dynamics. 
Figure~\ref{Fig.3}(b) shows the $H$ dependence of $(1/T_1T)_{H \perp c}$. 
As reported previously\cite{S.Kitagawa_PRL_2011}, at $x = 1$, the low-field PM state with the HF character is separated from the high-field polarized PM state with small $(1/T_1T)_{H \perp c}$ by a broad peak of $(1/T_1T)_{H \perp c}$ at $H_{\rm M} \simeq$ 4.3~T. Similar broad peak $(1/T_1T)_{H \perp c}$ was observed at $x = 0.95$, but is changed to the shoulder structure at $x = 0.9$. 
On the other hand, $(1/T_1T)_{H \perp c}$ is monotonically suppressed by magnetic field and no anomaly is observed at $x = 0.85$ and 0.87.
We determined $H_{\rm M}$ from the $H$ dependence of $M_{\perp}$ and $(1/T_1T)_{H \perp c}$ as shown by the dotted lines in Fig.~\ref{Fig.3}.
At $x = 0.85$, we could not observe any MM behavior both in $M_{\perp}$ and $(1/T_1T)_{H \perp c}$.

\begin{figure}[tb]
\vspace*{-10pt}
\begin{center}
\includegraphics[width=8.5cm,clip]{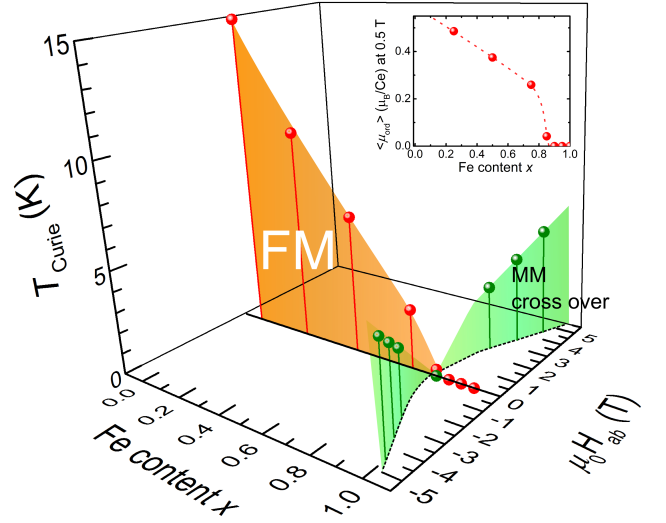}
\end{center}
\caption{(Color online)  $x-T-H$ phase diagram in Ce(Ru$_{1-x}$Fe$_{x}$)PO determined by NMR results. $T_{\rm C}$ in CeRuPO is continuously suppressed by Fe substitution, and approaches to zero at $x \sim 0.86$, suggestive of the presence of a FM QCP at $x \sim 0.86$. 
Furthermore, $H_{\rm M}$ observed in CeFePO decreases with decreasing $x$ and seems to merge with the FM QCP at $x \sim 0.86$. 
(Inset) $x$ dependence of ordered magnetic moment $\langle \mu_{\rm ord} \rangle$  estimated from $K_c$. $\langle \mu_{\rm ord} \rangle$ continuously approaches to zero toward $x \sim 0.86$, which is consistent with $x$ dependence of $T_{\rm C}$. Dashed curve is guide to eyes.
The 2nd-order FM transition is suggested from the development of the critical fluctuations down to low temperatures as well as the continuous suppression of $T_{\rm C}$ and the ordered moments $\langle \mu_{\rm ord} \rangle$. }
\label{Fig.4}
\end{figure}

On the basis of above results, we developed the $x-T-H$ phase diagram on Ce(Ru$_{1-x}$Fe$_{x}$)PO in Fig.~\ref{Fig.4}. $T_{\rm C}$ in CeRuPO is continuously suppressed by the Fe substitution, and approaches to zero at $x \sim 0.86$. 
This is consistent with the $x$ dependence of $\langle \mu_{\rm ord} \rangle$ evaluated from the $K_{\rm c}$, which is shown in the inset of Fig.~\ref{Fig.4}. 
In addition, remarkable divergence of $S_{\perp}$ against $T$ at $x = 0.85$ and against $x$ at 0.2~K as shown in Fig.~\ref{Fig.2} strongly suggest the presence of a FM QCP at $x \sim 0.86$ in Ce(Ru$_{1-x}$Fe$_{x}$)PO. 
Furthermore, $H_{\rm M}$ observed in CeFePO decreases with decreasing $x$ and seems to merge with the FM QCP at $x \sim 0.86$, suggestive of the close relationship between the FM QCP and the criticality of the MM behavior. 

There have been a lot of theoretical and experimental efforts for understanding a FM quantum criticality, since a 1st-order FM transition has been observed in various itinerant ferromagnets, e.g. ZrZn$_2$, MnSi, (Sr$_{1-x}$Ca$_{x}$)RuO$_3$ and so on\cite{Y.J.Uemura_NaturePhys_2007}.
In addition, Belitz $et~al$. proposed that the 1st-order MM transitions emerge from the TCP in the small field range and terminate at quantum critical end points\cite{D.Belitz_PRL_2005}. 
This characteristic phase diagram was experimentally confirmed in UGe$_2$ \cite{H.Kotegawa_JPSJ_2011}, URhGe \cite{F.Levy_NaturePhys_2007} and partly in UCoAl\cite{D.Aoki_JPSJ_2011}.
The phase diagram obtained in Ce(Ru$_{1-x}$Fe$_{x}$)PO is quite different from that in other itinerant ferromagnets reported so far [e.g. UGe$_2$ and Co(S$_{1-x}$Se$_x$)$_2$\cite{K.Adachi_JPSJ_1979}]. 
Particularly, the 2nd-order character of the FM QCP is a rare example among itinerant ferromagnets\cite{C.Krellner_NJP_2011}. 
However, it might be possible to reconcile the different phase diagram by the suppression of a TCP temperature to nearly zero or a negative value in Ce(Ru$_{1-x}$Fe$_{x}$)PO. 
Absence of a 1st-order MM transition is also consistent with the 2nd-order character of the FM-PM transition. 

\begin{figure}[tb]
\vspace*{-10pt}
\begin{center}
\includegraphics[width=11cm,clip]{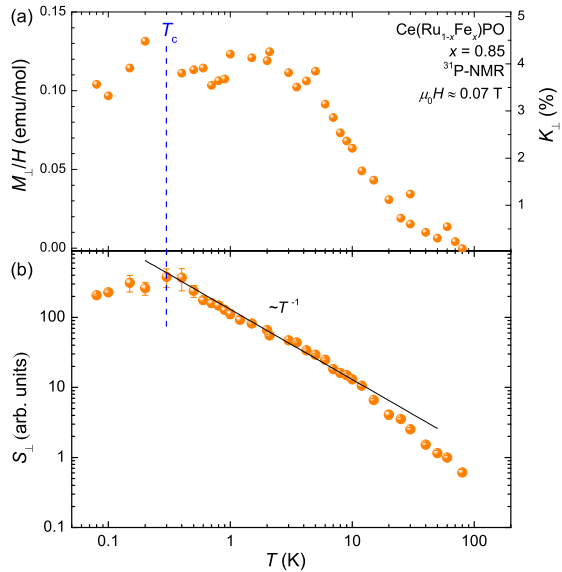}
\end{center}
\caption{(Color online)  $T$ dependence of $M_{\perp}/H$ ($K_{\perp}$) (a) and $S_{\perp}$ (b) measured at the smallest field we could ( $\mu_0H \simeq 0.07$~T) at $x = 0.85$. Static susceptibility related with $q = 0$ becomes saturate below 3~K, although $S_{\perp}$ continues to develop toward $T_{\rm C}$, suggesting that spin fluctuations would possess AFM ($q \neq  0$) components as well as the FM ($q = 0$) components near the FM QCP. These behaviors were in sharp contrast with those observed in $x < 0.75$ as shown in Fig.~\ref{Fig.1} (b), but were actually observed in YbRh$_2$Si$_2$ near an $H$-induced AFM QCP, in which the possibility of the local criticality has been discussed. }
\label{Fig.5}
\end{figure}

In addition to the novel 2nd-order FM QCP, it deserves to be noted that the ``wing'' of the MM crossover also converges at $x \sim 0.86$. 
Since the $H_{\rm M}$ is related to the characteristic energy of the HF state, this reminds us of the local QCP observed in YbRh$_2$Si$_2$\cite{S.Paschen_Nature_2004,Q.Si_Science_2010}. 
Actually, similarity was observed between critical behaviors in YbRh$_2$Si$_2$ and those in Ce(Ru$_{0.15}$Fe$_{0.85}$)PO.
Figure~\ref{Fig.5} shows the $T$ dependence of $M_{\perp}/H$ ($K_{\perp}$) and $S_{\perp}$ measured at the smallest field we could ( $\mu_0H \simeq 0.07$~T) at $x = 0.85$. 
Static susceptibility related with $q = 0$ becomes saturate below 3~K, although $S_{\perp}$ continues to develop with a $S_{\perp} \propto T^{-1}$ relation toward $T_{\rm C}$, suggesting that spin fluctuations would possess AFM ($q \neq 0$) components as well as the FM ($q = 0$) components near the FM QCP. 
These behaviors were in sharp contrast with those observed in $x < 0.75$ [see in Fig.~\ref{Fig.1}(b)], but were actually observed in YbRh$_2$Si$_2$ near AFM QCP\cite{K.Ishida_PRL_2002}. 
It is noted that both compounds possess the two-dimensional $XY$ anisotropy in the PM state, which might be related to the origin of the local QCP. 
For the local QCP, a sudden change of the Fermi surface is expected at the QCP, where magnetic fluctuations at various $q$ components are considered to develop due to the Fermi-surface instability.
Thus, we speculate that sudden Fermi-surface change might occur at $x \sim 0.86$, which cannot be understood by the ordinary spin density wave scenario\cite{J.A.Hertz_PRB_1976,A.J.Millis_PRB_1993,T.Moriya_JPSJ_1995}.

In summary, both of the static and dynamic susceptibilities in the FM and PM states of Ce(Ru$_{1-x}$Fe$_{x}$)PO clearly suggest that the FM QCP is located at $x \sim 0.86$, and that the FM QCP possesses a 2nd-order character, which is a rare example for itinerant ferromagnets. 
In addition, continuous suppression of the MM crossover toward $x \sim 0.86$, and the characteristic behavior of $S_{\perp}$ suggests that the close relationship between the FM QCP and the criticality of the Kondo effect, which can be understood by the local QCP scenario. 

The authors thank to Y. Nakai, K. Kitagawa, S. Yonezawa, and Y. Maeno for experimental support and valuable discussions. 
The authors also grateful to H. Ikeda for fruitful discussion. 
This work was partially supported by Kyoto Univ. LTM center, the ``Heavy Electrons'' Grant-in-Aid for Scientific Research on Innovative Areas  (No. 20102006) from The Ministry of Education, Culture, Sports, Science, and Technology (MEXT) of Japan, a Grant-in-Aid for the Global COE Program ``The Next Generation of Physics, Spun from Universality and Emergence'' from MEXT of Japan, a grant-in-aid for Scientific Research from Japan Society for Promotion of Science (JSPS), KAKENHI (S \& A) (No. 20224008 \& No. 23244075), Funding Program for World-Leading Innovative R\&D on Science and Technology (FIRST) from JSPS. One of the authors (SK) is financially supported by a JSPS Research Fellowship.


\begin{thebibliography}{31}%
\makeatletter
\providecommand \@ifxundefined [1]{%
 \@ifx{#1\undefined}
}%
\providecommand \@ifnum [1]{%
 \ifnum #1\expandafter \@firstoftwo
 \else \expandafter \@secondoftwo
 \fi
}%
\providecommand \@ifx [1]{%
 \ifx #1\expandafter \@firstoftwo
 \else \expandafter \@secondoftwo
 \fi
}%
\providecommand \natexlab [1]{#1}%
\providecommand \enquote  [1]{``#1''}%
\providecommand \bibnamefont  [1]{#1}%
\providecommand \bibfnamefont [1]{#1}%
\providecommand \citenamefont [1]{#1}%
\providecommand \href@noop [0]{\@secondoftwo}%
\providecommand \href [0]{\begingroup \@sanitize@url \@href}%
\providecommand \@href[1]{\@@startlink{#1}\@@href}%
\providecommand \@@href[1]{\endgroup#1\@@endlink}%
\providecommand \@sanitize@url [0]{\catcode `\\12\catcode `\$12\catcode
  `\&12\catcode `\#12\catcode `\^12\catcode `\_12\catcode `\%12\relax}%
\providecommand \@@startlink[1]{}%
\providecommand \@@endlink[0]{}%
\providecommand \url  [0]{\begingroup\@sanitize@url \@url }%
\providecommand \@url [1]{\endgroup\@href {#1}{\urlprefix }}%
\providecommand \urlprefix  [0]{URL }%
\providecommand \Eprint [0]{\href }%
\providecommand \doibase [0]{http://dx.doi.org/}%
\providecommand \selectlanguage [0]{\@gobble}%
\providecommand \bibinfo  [0]{\@secondoftwo}%
\providecommand \bibfield  [0]{\@secondoftwo}%
\providecommand \translation [1]{[#1]}%
\providecommand \BibitemOpen [0]{}%
\providecommand \bibitemStop [0]{}%
\providecommand \bibitemNoStop [0]{.\EOS\space}%
\providecommand \EOS [0]{\spacefactor3000\relax}%
\providecommand \BibitemShut  [1]{\csname bibitem#1\endcsname}%
\let\auto@bib@innerbib\@empty
\bibitem [{\citenamefont {Hertz}(1976)}]{J.A.Hertz_PRB_1976}%
  \BibitemOpen
  \bibfield  {author} {\bibinfo {author} {\bibfnamefont {J.~A.}\ \bibnamefont
  {Hertz}},\ }\href@noop {} {\bibfield  {journal} {\bibinfo  {journal} {Phys.
  Rev. B}\ }\textbf {\bibinfo {volume} {14}},\ \bibinfo {pages} {1165}
  (\bibinfo {year} {1976})}\BibitemShut {NoStop}%
\bibitem [{\citenamefont {Uhlarz}\ \emph {et~al.}(2004)\citenamefont {Uhlarz},
  \citenamefont {C.},\ and\ \citenamefont {Hayden}}]{M.Uhlarz_PRL_2004}%
  \BibitemOpen
  \bibfield  {author} {\bibinfo {author} {\bibfnamefont {M.}~\bibnamefont
  {Uhlarz}}, \bibinfo {author} {\bibfnamefont {P.}~\bibnamefont {C.}}, \ and\
  \bibinfo {author} {\bibfnamefont {S.~M.}\ \bibnamefont {Hayden}},\
  }\href@noop {} {\bibfield  {journal} {\bibinfo  {journal} {Phys. Rev. Lett.}\
  }\textbf {\bibinfo {volume} {93}},\ \bibinfo {pages} {256404} (\bibinfo
  {year} {2004})}\BibitemShut {NoStop}%
\bibitem [{\citenamefont {Pfleiderer}\ \emph {et~al.}(1997)\citenamefont
  {Pfleiderer}, \citenamefont {McMullan}, \citenamefont {J.~Julian},\ and\
  \citenamefont {Lonzarich}}]{C.Pfleiderer_PRB_1997}%
  \BibitemOpen
  \bibfield  {author} {\bibinfo {author} {\bibfnamefont {C.}~\bibnamefont
  {Pfleiderer}}, \bibinfo {author} {\bibfnamefont {G.}~\bibnamefont
  {McMullan}}, \bibinfo {author} {\bibfnamefont {S.~R.}\ \bibnamefont
  {J.~Julian}}, \ and\ \bibinfo {author} {\bibfnamefont {G.~G.}\ \bibnamefont
  {Lonzarich}},\ }\href@noop {} {\bibfield  {journal} {\bibinfo  {journal}
  {Phys. Rev. B}\ }\textbf {\bibinfo {volume} {55}},\ \bibinfo {pages} {8330}
  (\bibinfo {year} {1997})}\BibitemShut {NoStop}%
\bibitem [{\citenamefont {Levy}\ \emph {et~al.}(2007)\citenamefont {Levy},
  \citenamefont {Sheikin},\ and\ \citenamefont
  {Huxley}}]{F.Levy_NaturePhys_2007}%
  \BibitemOpen
  \bibfield  {author} {\bibinfo {author} {\bibfnamefont {F.}~\bibnamefont
  {Levy}}, \bibinfo {author} {\bibfnamefont {I.}~\bibnamefont {Sheikin}}, \
  and\ \bibinfo {author} {\bibfnamefont {A.}~\bibnamefont {Huxley}},\
  }\href@noop {} {\bibfield  {journal} {\bibinfo  {journal} {Nature Physics}\
  }\textbf {\bibinfo {volume} {3}},\ \bibinfo {pages} {460} (\bibinfo {year}
  {2007})}\BibitemShut {NoStop}%
\bibitem [{\citenamefont {Kotegawa}\ \emph {et~al.}(2011)\citenamefont
  {Kotegawa}, \citenamefont {Taufour}, \citenamefont {Aoki}, \citenamefont
  {Knebel},\ and\ \citenamefont {Flouquet}}]{H.Kotegawa_JPSJ_2011}%
  \BibitemOpen
  \bibfield  {author} {\bibinfo {author} {\bibfnamefont {H.}~\bibnamefont
  {Kotegawa}}, \bibinfo {author} {\bibfnamefont {V.}~\bibnamefont {Taufour}},
  \bibinfo {author} {\bibfnamefont {D.}~\bibnamefont {Aoki}}, \bibinfo {author}
  {\bibfnamefont {G.}~\bibnamefont {Knebel}}, \ and\ \bibinfo {author}
  {\bibfnamefont {J.}~\bibnamefont {Flouquet}},\ }\href@noop {} {\bibfield
  {journal} {\bibinfo  {journal} {J. Phys. Soc. Jpn.}\ }\textbf {\bibinfo
  {volume} {80}},\ \bibinfo {pages} {083703} (\bibinfo {year}
  {2011})}\BibitemShut {NoStop}%
\bibitem [{\citenamefont {Adachi}\ \emph {et~al.}(1979)\citenamefont {Adachi},
  \citenamefont {Matsui}, \citenamefont {Omata}, \citenamefont {Mollymoto},
  \citenamefont {Motokawa},\ and\ \citenamefont {Date}}]{K.Adachi_JPSJ_1979}%
  \BibitemOpen
  \bibfield  {author} {\bibinfo {author} {\bibfnamefont {K.}~\bibnamefont
  {Adachi}}, \bibinfo {author} {\bibfnamefont {M.}~\bibnamefont {Matsui}},
  \bibinfo {author} {\bibfnamefont {Y.}~\bibnamefont {Omata}}, \bibinfo
  {author} {\bibfnamefont {H.}~\bibnamefont {Mollymoto}}, \bibinfo {author}
  {\bibfnamefont {M.}~\bibnamefont {Motokawa}}, \ and\ \bibinfo {author}
  {\bibfnamefont {M.}~\bibnamefont {Date}},\ }\href@noop {} {\bibfield
  {journal} {\bibinfo  {journal} {J. Phys. Soc. Jpn.}\ }\textbf {\bibinfo
  {volume} {47}},\ \bibinfo {pages} {675} (\bibinfo {year} {1979})}\BibitemShut
  {NoStop}%
\bibitem [{\citenamefont {Yamada}(1993)}]{H.Yamada_PRB_1993}%
  \BibitemOpen
  \bibfield  {author} {\bibinfo {author} {\bibfnamefont {H.}~\bibnamefont
  {Yamada}},\ }\href@noop {} {\bibfield  {journal} {\bibinfo  {journal} {Phys.
  Rev. B}\ }\textbf {\bibinfo {volume} {47}},\ \bibinfo {pages} {11211}
  (\bibinfo {year} {1993})}\BibitemShut {NoStop}%
\bibitem [{\citenamefont {Belitz}\ \emph {et~al.}(2005)\citenamefont {Belitz},
  \citenamefont {Kirkpatrick},\ and\ \citenamefont
  {Rollbuhler}}]{D.Belitz_PRL_2005}%
  \BibitemOpen
  \bibfield  {author} {\bibinfo {author} {\bibfnamefont {D.}~\bibnamefont
  {Belitz}}, \bibinfo {author} {\bibfnamefont {T.~R.}\ \bibnamefont
  {Kirkpatrick}}, \ and\ \bibinfo {author} {\bibfnamefont {J.}~\bibnamefont
  {Rollbuhler}},\ }\href@noop {} {\bibfield  {journal} {\bibinfo  {journal}
  {Phys. Rev. Lett.}\ }\textbf {\bibinfo {volume} {94}},\ \bibinfo {pages}
  {247205} (\bibinfo {year} {2005})}\BibitemShut {NoStop}%
\bibitem [{\citenamefont {Knebel}\ \emph {et~al.}(2006)\citenamefont {Knebel},
  \citenamefont {Aoki}, \citenamefont {Braithwaite}, \citenamefont {Salce},\
  and\ \citenamefont {Flouquet}}]{G.Knebel_PRB_2006}%
  \BibitemOpen
  \bibfield  {author} {\bibinfo {author} {\bibfnamefont {G.}~\bibnamefont
  {Knebel}}, \bibinfo {author} {\bibfnamefont {D.}~\bibnamefont {Aoki}},
  \bibinfo {author} {\bibfnamefont {D.}~\bibnamefont {Braithwaite}}, \bibinfo
  {author} {\bibfnamefont {B.}~\bibnamefont {Salce}}, \ and\ \bibinfo {author}
  {\bibfnamefont {J.}~\bibnamefont {Flouquet}},\ }\href@noop {} {\bibfield
  {journal} {\bibinfo  {journal} {Phys. Rev. B}\ }\textbf {\bibinfo {volume}
  {74}},\ \bibinfo {pages} {020501(R)} (\bibinfo {year} {2006})}\BibitemShut
  {NoStop}%
\bibitem [{\citenamefont {Kasahara}\ \emph {et~al.}(2010)\citenamefont
  {Kasahara}, \citenamefont {Shibauchi}, \citenamefont {Hashimoto},
  \citenamefont {Ikada}, \citenamefont {Tonegawa}, \citenamefont {Okazaki},
  \citenamefont {Shishido}, \citenamefont {Ikeda}, \citenamefont {Takeya},
  \citenamefont {Hirata}, \citenamefont {Terashima},\ and\ \citenamefont
  {Matsuda}}]{S.Kasahara_PRB_2010}%
  \BibitemOpen
  \bibfield  {author} {\bibinfo {author} {\bibfnamefont {S.}~\bibnamefont
  {Kasahara}}, \bibinfo {author} {\bibfnamefont {T.}~\bibnamefont {Shibauchi}},
  \bibinfo {author} {\bibfnamefont {K.}~\bibnamefont {Hashimoto}}, \bibinfo
  {author} {\bibfnamefont {K.}~\bibnamefont {Ikada}}, \bibinfo {author}
  {\bibfnamefont {S.}~\bibnamefont {Tonegawa}}, \bibinfo {author}
  {\bibfnamefont {R.}~\bibnamefont {Okazaki}}, \bibinfo {author} {\bibfnamefont
  {H.}~\bibnamefont {Shishido}}, \bibinfo {author} {\bibfnamefont
  {H.}~\bibnamefont {Ikeda}}, \bibinfo {author} {\bibfnamefont
  {H.}~\bibnamefont {Takeya}}, \bibinfo {author} {\bibfnamefont
  {K.}~\bibnamefont {Hirata}}, \bibinfo {author} {\bibfnamefont
  {T.}~\bibnamefont {Terashima}}, \ and\ \bibinfo {author} {\bibfnamefont
  {Y.}~\bibnamefont {Matsuda}},\ }\href@noop {} {\bibfield  {journal} {\bibinfo
   {journal} {Phys. Rev. B}\ }\textbf {\bibinfo {volume} {81}},\ \bibinfo
  {pages} {184519} (\bibinfo {year} {2010})}\BibitemShut {NoStop}%
\bibitem [{\citenamefont {Borzi}\ \emph {et~al.}(2007)\citenamefont {Borzi},
  \citenamefont {Grigera}, \citenamefont {Farrell}, \citenamefont {Perry},
  \citenamefont {Lister}, \citenamefont {Lee}, \citenamefont {Tennant},
  \citenamefont {Maeno},\ and\ \citenamefont
  {Mackenzie}}]{R.A.Borzi_Science_2007}%
  \BibitemOpen
  \bibfield  {author} {\bibinfo {author} {\bibfnamefont {R.~A.}\ \bibnamefont
  {Borzi}}, \bibinfo {author} {\bibfnamefont {S.~A.}\ \bibnamefont {Grigera}},
  \bibinfo {author} {\bibfnamefont {J.}~\bibnamefont {Farrell}}, \bibinfo
  {author} {\bibfnamefont {R.~S.}\ \bibnamefont {Perry}}, \bibinfo {author}
  {\bibfnamefont {S.~J.~S.}\ \bibnamefont {Lister}}, \bibinfo {author}
  {\bibfnamefont {S.~L.}\ \bibnamefont {Lee}}, \bibinfo {author} {\bibfnamefont
  {D.~A.}\ \bibnamefont {Tennant}}, \bibinfo {author} {\bibfnamefont
  {Y.}~\bibnamefont {Maeno}}, \ and\ \bibinfo {author} {\bibfnamefont {A.~P.}\
  \bibnamefont {Mackenzie}},\ }\href@noop {} {\bibfield  {journal} {\bibinfo
  {journal} {Science}\ }\textbf {\bibinfo {volume} {315}},\ \bibinfo {pages}
  {214} (\bibinfo {year} {2007})}\BibitemShut {NoStop}%
\bibitem [{\citenamefont {Schroder}\ \emph {et~al.}(2000)\citenamefont
  {Schroder}, \citenamefont {Aeppli}, \citenamefont {Coldea}, \citenamefont
  {Adams}, \citenamefont {Stockert}, \citenamefont {Lohneysen}, \citenamefont
  {Bucher}, \citenamefont {Ramazashvili},\ and\ \citenamefont
  {Coleman}}]{A.Schroder_Nature_2000}%
  \BibitemOpen
  \bibfield  {author} {\bibinfo {author} {\bibfnamefont {A.}~\bibnamefont
  {Schroder}}, \bibinfo {author} {\bibfnamefont {G.}~\bibnamefont {Aeppli}},
  \bibinfo {author} {\bibfnamefont {R.}~\bibnamefont {Coldea}}, \bibinfo
  {author} {\bibfnamefont {M.}~\bibnamefont {Adams}}, \bibinfo {author}
  {\bibfnamefont {O.}~\bibnamefont {Stockert}}, \bibinfo {author}
  {\bibfnamefont {H.}~\bibnamefont {Lohneysen}}, \bibinfo {author}
  {\bibfnamefont {E.}~\bibnamefont {Bucher}}, \bibinfo {author} {\bibfnamefont
  {R.}~\bibnamefont {Ramazashvili}}, \ and\ \bibinfo {author} {\bibfnamefont
  {P.}~\bibnamefont {Coleman}},\ }\href@noop {} {\bibfield  {journal} {\bibinfo
   {journal} {Nature}\ }\textbf {\bibinfo {volume} {407}},\ \bibinfo {pages}
  {351} (\bibinfo {year} {2000})}\BibitemShut {NoStop}%
\bibitem [{\citenamefont {Pagliuso}\ \emph {et~al.}(2001)\citenamefont
  {Pagliuso}, \citenamefont {Petrovic}, \citenamefont {Movshovich},
  \citenamefont {Hall}, \citenamefont {Hundley}, \citenamefont {Sarrao},
  \citenamefont {Thompson}, ,\ and\ \citenamefont
  {Fisk}}]{P.G.Pagliuso_PRB_2001}%
  \BibitemOpen
  \bibfield  {author} {\bibinfo {author} {\bibfnamefont {P.~G.}\ \bibnamefont
  {Pagliuso}}, \bibinfo {author} {\bibfnamefont {C.}~\bibnamefont {Petrovic}},
  \bibinfo {author} {\bibfnamefont {R.}~\bibnamefont {Movshovich}}, \bibinfo
  {author} {\bibfnamefont {D.}~\bibnamefont {Hall}}, \bibinfo {author}
  {\bibfnamefont {M.~F.}\ \bibnamefont {Hundley}}, \bibinfo {author}
  {\bibfnamefont {J.~L.}\ \bibnamefont {Sarrao}}, \bibinfo {author}
  {\bibfnamefont {J.~D.}\ \bibnamefont {Thompson}}, , \ and\ \bibinfo {author}
  {\bibfnamefont {Z.}~\bibnamefont {Fisk}},\ }\href@noop {} {\bibfield
  {journal} {\bibinfo  {journal} {Phys. Rev. B}\ }\textbf {\bibinfo {volume}
  {64}},\ \bibinfo {pages} {100503(R)} (\bibinfo {year} {2001})}\BibitemShut
  {NoStop}%
\bibitem [{\citenamefont {Ohashi}\ \emph {et~al.}(2003)\citenamefont {Ohashi},
  \citenamefont {Oomi}, \citenamefont {Koiwai}, \citenamefont {Hedo},\ and\
  \citenamefont {Uwatoko}}]{M.Ohashi_PRB_2003}%
  \BibitemOpen
  \bibfield  {author} {\bibinfo {author} {\bibfnamefont {M.}~\bibnamefont
  {Ohashi}}, \bibinfo {author} {\bibfnamefont {G.}~\bibnamefont {Oomi}},
  \bibinfo {author} {\bibfnamefont {S.}~\bibnamefont {Koiwai}}, \bibinfo
  {author} {\bibfnamefont {M.}~\bibnamefont {Hedo}}, \ and\ \bibinfo {author}
  {\bibfnamefont {Y.}~\bibnamefont {Uwatoko}},\ }\href@noop {} {\bibfield
  {journal} {\bibinfo  {journal} {Phys. Rev. B}\ }\textbf {\bibinfo {volume}
  {68}},\ \bibinfo {pages} {144428} (\bibinfo {year} {2003})}\BibitemShut
  {NoStop}%
\bibitem [{\citenamefont {Hassinger}\ \emph {et~al.}(2008)\citenamefont
  {Hassinger}, \citenamefont {Aoki}, \citenamefont {Knebel},\ and\
  \citenamefont {Flouquet}}]{E.Hassinger_JPSJ_2008}%
  \BibitemOpen
  \bibfield  {author} {\bibinfo {author} {\bibfnamefont {E.}~\bibnamefont
  {Hassinger}}, \bibinfo {author} {\bibfnamefont {D.}~\bibnamefont {Aoki}},
  \bibinfo {author} {\bibfnamefont {G.}~\bibnamefont {Knebel}}, \ and\ \bibinfo
  {author} {\bibfnamefont {J.}~\bibnamefont {Flouquet}},\ }\href@noop {}
  {\bibfield  {journal} {\bibinfo  {journal} {J. Phys. Soc. Jpn.}\ }\textbf
  {\bibinfo {volume} {77}},\ \bibinfo {pages} {073703} (\bibinfo {year}
  {2008})}\BibitemShut {NoStop}%
\bibitem [{\citenamefont {Gegenwart}\ \emph {et~al.}(2002)\citenamefont
  {Gegenwart}, \citenamefont {Custers}, \citenamefont {Geibel}, \citenamefont
  {Neumaier}, \citenamefont {Tayama}, \citenamefont {Tenya}, \citenamefont
  {Trovarelli},\ and\ \citenamefont {Steglich}}]{P.Gegenwart_PRL_2002}%
  \BibitemOpen
  \bibfield  {author} {\bibinfo {author} {\bibfnamefont {P.}~\bibnamefont
  {Gegenwart}}, \bibinfo {author} {\bibfnamefont {J.}~\bibnamefont {Custers}},
  \bibinfo {author} {\bibfnamefont {C.}~\bibnamefont {Geibel}}, \bibinfo
  {author} {\bibfnamefont {K.}~\bibnamefont {Neumaier}}, \bibinfo {author}
  {\bibfnamefont {T.}~\bibnamefont {Tayama}}, \bibinfo {author} {\bibfnamefont
  {K.}~\bibnamefont {Tenya}}, \bibinfo {author} {\bibfnamefont
  {O.}~\bibnamefont {Trovarelli}}, \ and\ \bibinfo {author} {\bibfnamefont
  {F.}~\bibnamefont {Steglich}},\ }\href@noop {} {\bibfield  {journal}
  {\bibinfo  {journal} {Phys. Rev. Lett.}\ }\textbf {\bibinfo {volume} {89}},\
  \bibinfo {pages} {056402} (\bibinfo {year} {2002})}\BibitemShut {NoStop}%
\bibitem [{\citenamefont {Kamihara}\ \emph {et~al.}(2006)\citenamefont
  {Kamihara}, \citenamefont {Hiramatsu}, \citenamefont {Hirano}, \citenamefont
  {Kawamura}, \citenamefont {Yanagi}, \citenamefont {Kamiya},\ and\
  \citenamefont {Hosono}}]{Y.Kamihara_JACS_2006}%
  \BibitemOpen
  \bibfield  {author} {\bibinfo {author} {\bibfnamefont {Y.}~\bibnamefont
  {Kamihara}}, \bibinfo {author} {\bibfnamefont {H.}~\bibnamefont {Hiramatsu}},
  \bibinfo {author} {\bibfnamefont {M.}~\bibnamefont {Hirano}}, \bibinfo
  {author} {\bibfnamefont {R.}~\bibnamefont {Kawamura}}, \bibinfo {author}
  {\bibfnamefont {H.}~\bibnamefont {Yanagi}}, \bibinfo {author} {\bibfnamefont
  {T.}~\bibnamefont {Kamiya}}, \ and\ \bibinfo {author} {\bibfnamefont
  {H.}~\bibnamefont {Hosono}},\ }\href@noop {} {\bibfield  {journal} {\bibinfo
  {journal} {J. Am. Chem. Soc.}\ }\textbf {\bibinfo {volume} {128}},\ \bibinfo
  {pages} {10012} (\bibinfo {year} {2006})}\BibitemShut {NoStop}%
\bibitem [{\citenamefont {Krellner}\ \emph {et~al.}(2007)\citenamefont
  {Krellner}, \citenamefont {Kini}, \citenamefont {Bruning}, \citenamefont
  {Koch}, \citenamefont {Rosner}, \citenamefont {Nicklas}, \citenamefont
  {Baenitz},\ and\ \citenamefont {Geibel}}]{C.Krellner_PRB_2007}%
  \BibitemOpen
  \bibfield  {author} {\bibinfo {author} {\bibfnamefont {C.}~\bibnamefont
  {Krellner}}, \bibinfo {author} {\bibfnamefont {N.~S.}\ \bibnamefont {Kini}},
  \bibinfo {author} {\bibfnamefont {E.~M.}\ \bibnamefont {Bruning}}, \bibinfo
  {author} {\bibfnamefont {K.}~\bibnamefont {Koch}}, \bibinfo {author}
  {\bibfnamefont {H.}~\bibnamefont {Rosner}}, \bibinfo {author} {\bibfnamefont
  {M.}~\bibnamefont {Nicklas}}, \bibinfo {author} {\bibfnamefont
  {M.}~\bibnamefont {Baenitz}}, \ and\ \bibinfo {author} {\bibfnamefont
  {C.}~\bibnamefont {Geibel}},\ }\href@noop {} {\bibfield  {journal} {\bibinfo
  {journal} {Phys. Rev. B}\ }\textbf {\bibinfo {volume} {76}},\ \bibinfo
  {pages} {104418} (\bibinfo {year} {2007})}\BibitemShut {NoStop}%
\bibitem [{\citenamefont {Bruning}\ \emph {et~al.}(2008)\citenamefont
  {Bruning}, \citenamefont {Krellner}, \citenamefont {Baenitz}, \citenamefont
  {Jesche}, \citenamefont {Steglich},\ and\ \citenamefont
  {Geibel}}]{E.Buning_PRL_2008}%
  \BibitemOpen
  \bibfield  {author} {\bibinfo {author} {\bibfnamefont {E.~M.}\ \bibnamefont
  {Bruning}}, \bibinfo {author} {\bibfnamefont {C.}~\bibnamefont {Krellner}},
  \bibinfo {author} {\bibfnamefont {M.}~\bibnamefont {Baenitz}}, \bibinfo
  {author} {\bibfnamefont {A.}~\bibnamefont {Jesche}}, \bibinfo {author}
  {\bibfnamefont {F.}~\bibnamefont {Steglich}}, \ and\ \bibinfo {author}
  {\bibfnamefont {C.}~\bibnamefont {Geibel}},\ }\href@noop {} {\bibfield
  {journal} {\bibinfo  {journal} {Phys. Rev. Lett.}\ }\textbf {\bibinfo
  {volume} {101}},\ \bibinfo {pages} {117206} (\bibinfo {year}
  {2008})}\BibitemShut {NoStop}%
\bibitem [{\citenamefont {Kitagawa}\ \emph {et~al.}(2011)\citenamefont
  {Kitagawa}, \citenamefont {Ikeda}, \citenamefont {Nakai}, \citenamefont
  {Hattori}, \citenamefont {Ishida}, \citenamefont {Kamihara}, \citenamefont
  {Hirano},\ and\ \citenamefont {Hosono}}]{S.Kitagawa_PRL_2011}%
  \BibitemOpen
  \bibfield  {author} {\bibinfo {author} {\bibfnamefont {S.}~\bibnamefont
  {Kitagawa}}, \bibinfo {author} {\bibfnamefont {H.}~\bibnamefont {Ikeda}},
  \bibinfo {author} {\bibfnamefont {Y.}~\bibnamefont {Nakai}}, \bibinfo
  {author} {\bibfnamefont {T.}~\bibnamefont {Hattori}}, \bibinfo {author}
  {\bibfnamefont {K.}~\bibnamefont {Ishida}}, \bibinfo {author} {\bibfnamefont
  {Y.}~\bibnamefont {Kamihara}}, \bibinfo {author} {\bibfnamefont
  {M.}~\bibnamefont {Hirano}}, \ and\ \bibinfo {author} {\bibfnamefont
  {H.}~\bibnamefont {Hosono}},\ }\href@noop {} {\bibfield  {journal} {\bibinfo
  {journal} {Phys. Rev. Lett.}\ }\textbf {\bibinfo {volume} {107}},\ \bibinfo
  {pages} {277002} (\bibinfo {year} {2011})}\BibitemShut {NoStop}%
\bibitem [{\citenamefont {Luo}\ \emph {et~al.}(2012)\citenamefont {Luo},
  \citenamefont {}, \citenamefont {Jiang}, \citenamefont {Dai}, \citenamefont
  {Cao},\ and\ \citenamefont {Xu}}]{Y.Luo_PRB_2010}%
  \BibitemOpen
  \bibfield  {author} {\bibinfo {author} {\bibfnamefont {Y.}~\bibnamefont
  {Luo}}, \bibinfo {author} {\bibfnamefont {Y.}~\bibnamefont {Li}}, \bibinfo
  {author} {\bibfnamefont {S.}~\bibnamefont {Jiang}}, \bibinfo {author}
  {\bibfnamefont {J.}~\bibnamefont {Dai}}, \bibinfo {author} {\bibfnamefont
  {G.}~\bibnamefont {Cao}}, \ and\ \bibinfo {author} {\bibfnamefont {Z.-a.}\
  \bibnamefont {Xu}},\ }\href@noop {} {\bibfield  {journal} {\bibinfo
  {journal} {Phys. Rev. B}\ }\textbf {\bibinfo {volume} {81}},\ \bibinfo
  {pages} {134422} (\bibinfo {year} {2010})}\BibitemShut {NoStop}%
\bibitem [{\citenamefont {Karube}\ \emph {et~al.}(2010)\citenamefont {Karube},
  \citenamefont {Hattori}, \citenamefont {Kitagawa}, \citenamefont {Ishida}, \citenamefont
  {Kimura},\ and\ \citenamefont {Komatsubara}}]{K.Karube_PRB_2012}%
  \BibitemOpen
  \bibfield  {author} {\bibinfo {author} {\bibfnamefont {K.}~\bibnamefont
  {Karube}}, \bibinfo {author} {\bibfnamefont {T.}~\bibnamefont {Hattori}}, \bibinfo
  {author} {\bibfnamefont {S.}~\bibnamefont {Kitagawa}}, \bibinfo {author}
  {\bibfnamefont {K.}~\bibnamefont {Ishida}}, \bibinfo {author} {\bibfnamefont
  {N.}~\bibnamefont {Kimura}}, \ and\ \bibinfo {author} {\bibfnamefont {T.}\
  \bibnamefont {Komatsubara}},\ }\href@noop {} {\bibfield  {journal} {\bibinfo
  {journal} {Phys. Rev. B}\ }\textbf {\bibinfo {volume} {86}},\ \bibinfo
  {pages} {024428} (\bibinfo {year} {2012})}\BibitemShut {NoStop}%
\bibitem [{\citenamefont {Kamihara}\ \emph {et~al.}(2008)\citenamefont
  {Kamihara}, \citenamefont {Hiramatsu}, \citenamefont {Hirano}, \citenamefont
  {Yanagi}, \citenamefont {Kamiya},\ and\ \citenamefont
  {Hosono}}]{Y.Kamihara_JPCS_2008}%
  \BibitemOpen
  \bibfield  {author} {\bibinfo {author} {\bibfnamefont {Y.}~\bibnamefont
  {Kamihara}}, \bibinfo {author} {\bibfnamefont {H.}~\bibnamefont {Hiramatsu}},
  \bibinfo {author} {\bibfnamefont {M.}~\bibnamefont {Hirano}}, \bibinfo
  {author} {\bibfnamefont {H.}~\bibnamefont {Yanagi}}, \bibinfo {author}
  {\bibfnamefont {T.}~\bibnamefont {Kamiya}}, \ and\ \bibinfo {author}
  {\bibfnamefont {H.}~\bibnamefont {Hosono}},\ }\href@noop {} {\bibfield
  {journal} {\bibinfo  {journal} {J. Phys. Chem. Sol.}\ }\textbf {\bibinfo
  {volume} {69}},\ \bibinfo {pages} {2916} (\bibinfo {year}
  {2008})}\BibitemShut {NoStop}%
\bibitem [{\citenamefont {Nakamura}\ \emph {et~al.}()\citenamefont {Nakamura},
  \citenamefont {Iritani}, \citenamefont {Yano}, \citenamefont {Matoba},\ and\
  \citenamefont {Kamihara}}]{T.Nakamura_ICM_2012}%
  \BibitemOpen
  \bibfield  {author} {\bibinfo {author} {\bibfnamefont {T.}~\bibnamefont
  {Nakamura}}, \bibinfo {author} {\bibfnamefont {K.}~\bibnamefont {Iritani}},
  \bibinfo {author} {\bibfnamefont {R.}~\bibnamefont {Yano}}, \bibinfo {author}
  {\bibfnamefont {M.}~\bibnamefont {Matoba}}, \ and\ \bibinfo {author}
  {\bibfnamefont {Y.}~\bibnamefont {Kamihara}},\ }\href@noop {} {\bibinfo
  {journal} {submitted to Journal of the Korean Physical Society as proceedings
  for ICM2012}\ }\BibitemShut {NoStop}%
\bibitem [{\citenamefont {Uemura}\ \emph {et~al.}(2007)\citenamefont {Uemura},
  \citenamefont {Goko}, \citenamefont {Gat-Malureanu}, \citenamefont {Carlo},
  \citenamefont {Russo}, \citenamefont {Savici}, \citenamefont {Aczel},
  \citenamefont {MacDougall}, \citenamefont {Rodriguez}, \citenamefont {Luke},
  \citenamefont {Dunsiger}, \citenamefont {McCollam}, \citenamefont {Arai},
  \citenamefont {Pfleiderer}, \citenamefont {Boni}, \citenamefont {Yoshimura},
  \citenamefont {Baggio-Saitovitch}, \citenamefont {Fontes}, \citenamefont
  {Larrea}, \citenamefont {Sushko},\ and\ \citenamefont
  {Sereni}}]{Y.J.Uemura_NaturePhys_2007}%
  \BibitemOpen
\bibfield  {journal} {  }\bibfield  {author} {\bibinfo {author} {\bibfnamefont
  {Y.~J.}\ \bibnamefont {Uemura}}, \bibinfo {author} {\bibfnamefont
  {T.}~\bibnamefont {Goko}}, \bibinfo {author} {\bibfnamefont {I.~M.}\
  \bibnamefont {Gat-Malureanu}}, \bibinfo {author} {\bibfnamefont {J.~P.}\
  \bibnamefont {Carlo}}, \bibinfo {author} {\bibfnamefont {P.~L.}\ \bibnamefont
  {Russo}}, \bibinfo {author} {\bibfnamefont {A.~T.}\ \bibnamefont {Savici}},
  \bibinfo {author} {\bibfnamefont {A.}~\bibnamefont {Aczel}}, \bibinfo
  {author} {\bibfnamefont {G.~J.}\ \bibnamefont {MacDougall}}, \bibinfo
  {author} {\bibfnamefont {J.~A.}\ \bibnamefont {Rodriguez}}, \bibinfo {author}
  {\bibfnamefont {G.~M.}\ \bibnamefont {Luke}}, \bibinfo {author}
  {\bibfnamefont {S.~R.}\ \bibnamefont {Dunsiger}}, \bibinfo {author}
  {\bibfnamefont {A.}~\bibnamefont {McCollam}}, \bibinfo {author}
  {\bibfnamefont {J.}~\bibnamefont {Arai}}, \bibinfo {author} {\bibfnamefont
  {C.}~\bibnamefont {Pfleiderer}}, \bibinfo {author} {\bibfnamefont
  {P.}~\bibnamefont {Boni}}, \bibinfo {author} {\bibfnamefont {K.}~\bibnamefont
  {Yoshimura}}, \bibinfo {author} {\bibfnamefont {E.}~\bibnamefont
  {Baggio-Saitovitch}}, \bibinfo {author} {\bibfnamefont {M.~B.}\ \bibnamefont
  {Fontes}}, \bibinfo {author} {\bibfnamefont {J.}~\bibnamefont {Larrea}},
  \bibinfo {author} {\bibfnamefont {Y.~V.}\ \bibnamefont {Sushko}}, \ and\
  \bibinfo {author} {\bibfnamefont {J.}~\bibnamefont {Sereni}},\ }\href@noop {}
  {\bibfield  {journal} {\bibinfo  {journal} {Nature Physics}\ }\textbf
  {\bibinfo {volume} {3}},\ \bibinfo {pages} {29} (\bibinfo {year}
  {2007})}\BibitemShut {NoStop}%
\bibitem [{\citenamefont {Aoki}\ \emph {et~al.}(2011)\citenamefont {Aoki},
  \citenamefont {Combier}, \citenamefont {Taufour}, \citenamefont {Matsuda},
  \citenamefont {Knebel}, \citenamefont {Kotegawa},\ and\ \citenamefont
  {Flouquet}}]{D.Aoki_JPSJ_2011}%
  \BibitemOpen
  \bibfield  {author} {\bibinfo {author} {\bibfnamefont {D.}~\bibnamefont
  {Aoki}}, \bibinfo {author} {\bibfnamefont {T.}~\bibnamefont {Combier}},
  \bibinfo {author} {\bibfnamefont {V.}~\bibnamefont {Taufour}}, \bibinfo
  {author} {\bibfnamefont {T.~D.}\ \bibnamefont {Matsuda}}, \bibinfo {author}
  {\bibfnamefont {G.}~\bibnamefont {Knebel}}, \bibinfo {author} {\bibfnamefont
  {H.}~\bibnamefont {Kotegawa}}, \ and\ \bibinfo {author} {\bibfnamefont
  {J.}~\bibnamefont {Flouquet}},\ }\href@noop {} {\bibfield  {journal}
  {\bibinfo  {journal} {J. Phys. Soc. Jpn.}\ }\textbf {\bibinfo {volume}
  {80}},\ \bibinfo {pages} {094711} (\bibinfo {year} {2011})}\BibitemShut
  {NoStop}%
\bibitem [{\citenamefont {Krellner}\ \emph {et~al.}(2011)\citenamefont
  {Krellner}, \citenamefont {Lausberg}, \citenamefont {Steppke}, \citenamefont
  {Brando}, \citenamefont {Pedrero}, \citenamefont {Pfau}, \citenamefont
  {Tence}, \citenamefont {Rosner}, \citenamefont {Steglich},\ and\
  \citenamefont {Geibel}}]{C.Krellner_NJP_2011}%
  \BibitemOpen
  \bibfield  {author} {\bibinfo {author} {\bibfnamefont {C.}~\bibnamefont
  {Krellner}}, \bibinfo {author} {\bibfnamefont {S.}~\bibnamefont {Lausberg}},
  \bibinfo {author} {\bibfnamefont {A.}~\bibnamefont {Steppke}}, \bibinfo
  {author} {\bibfnamefont {M.}~\bibnamefont {Brando}}, \bibinfo {author}
  {\bibfnamefont {L.}~\bibnamefont {Pedrero}}, \bibinfo {author} {\bibfnamefont
  {H.}~\bibnamefont {Pfau}}, \bibinfo {author} {\bibfnamefont {S.}~\bibnamefont
  {Tence}}, \bibinfo {author} {\bibfnamefont {H.}~\bibnamefont {Rosner}},
  \bibinfo {author} {\bibfnamefont {F.}~\bibnamefont {Steglich}}, \ and\
  \bibinfo {author} {\bibfnamefont {C.}~\bibnamefont {Geibel}},\ }\href@noop {}
  {\bibfield  {journal} {\bibinfo  {journal} {New J. Phys.}\ }\textbf {\bibinfo
  {volume} {13}},\ \bibinfo {pages} {103014} (\bibinfo {year}
  {2011})}\BibitemShut {NoStop}%
\bibitem [{\citenamefont {Paschen}\ \emph {et~al.}(2004)\citenamefont
  {Paschen}, \citenamefont {Luhmann}, \citenamefont {Wirth}, \citenamefont
  {Gegenwart}, \citenamefont {Trovarelli}, \citenamefont {Geibel},
  \citenamefont {Steglich}, \citenamefont {Coleman},\ and\ \citenamefont
  {Si}}]{S.Paschen_Nature_2004}%
  \BibitemOpen
  \bibfield  {author} {\bibinfo {author} {\bibfnamefont {S.}~\bibnamefont
  {Paschen}}, \bibinfo {author} {\bibfnamefont {T.}~\bibnamefont {Luhmann}},
  \bibinfo {author} {\bibfnamefont {S.}~\bibnamefont {Wirth}}, \bibinfo
  {author} {\bibfnamefont {P.}~\bibnamefont {Gegenwart}}, \bibinfo {author}
  {\bibfnamefont {O.}~\bibnamefont {Trovarelli}}, \bibinfo {author}
  {\bibfnamefont {C.}~\bibnamefont {Geibel}}, \bibinfo {author} {\bibfnamefont
  {F.}~\bibnamefont {Steglich}}, \bibinfo {author} {\bibfnamefont
  {P.}~\bibnamefont {Coleman}}, \ and\ \bibinfo {author} {\bibfnamefont
  {Q.}~\bibnamefont {Si}},\ }\href@noop {} {\bibfield  {journal} {\bibinfo
  {journal} {Nature}\ }\textbf {\bibinfo {volume} {432}},\ \bibinfo {pages}
  {881} (\bibinfo {year} {2004})}\BibitemShut {NoStop}%
\bibitem [{\citenamefont {Si}\ and\ \citenamefont
  {Steglich}(2010)}]{Q.Si_Science_2010}%
  \BibitemOpen
  \bibfield  {author} {\bibinfo {author} {\bibfnamefont {Q.}~\bibnamefont
  {Si}}\ and\ \bibinfo {author} {\bibfnamefont {F.}~\bibnamefont {Steglich}},\
  }\href@noop {} {\bibfield  {journal} {\bibinfo  {journal} {Science}\ }\textbf
  {\bibinfo {volume} {329}},\ \bibinfo {pages} {1161} (\bibinfo {year}
  {2010})}\BibitemShut {NoStop}%
\bibitem [{\citenamefont {Ishida}\ \emph {et~al.}(2002)\citenamefont {Ishida},
  \citenamefont {Okamoto}, \citenamefont {Kawasaki}, \citenamefont {Kitaoka},
  \citenamefont {Trovarelli}, \citenamefont {Geibel},\ and\ \citenamefont
  {Steglich}}]{K.Ishida_PRL_2002}%
  \BibitemOpen
  \bibfield  {author} {\bibinfo {author} {\bibfnamefont {K.}~\bibnamefont
  {Ishida}}, \bibinfo {author} {\bibfnamefont {K.}~\bibnamefont {Okamoto}},
  \bibinfo {author} {\bibfnamefont {Y.}~\bibnamefont {Kawasaki}}, \bibinfo
  {author} {\bibfnamefont {Y.}~\bibnamefont {Kitaoka}}, \bibinfo {author}
  {\bibfnamefont {O.}~\bibnamefont {Trovarelli}}, \bibinfo {author}
  {\bibfnamefont {C.}~\bibnamefont {Geibel}}, \ and\ \bibinfo {author}
  {\bibfnamefont {F.}~\bibnamefont {Steglich}},\ }\href@noop {} {\bibfield
  {journal} {\bibinfo  {journal} {Phys. Rev. Lett.}\ }\textbf {\bibinfo
  {volume} {89}},\ \bibinfo {pages} {107202} (\bibinfo {year}
  {2002})}\BibitemShut {NoStop}%
\bibitem [{\citenamefont {Millis}(1993)}]{A.J.Millis_PRB_1993}%
  \BibitemOpen
  \bibfield  {author} {\bibinfo {author} {\bibfnamefont {A.~J.}\ \bibnamefont
  {Millis}},\ }\href@noop {} {\bibfield  {journal} {\bibinfo  {journal} {Phys.
  Rev. B}\ }\textbf {\bibinfo {volume} {48}},\ \bibinfo {pages} {7183}
  (\bibinfo {year} {1993})}\BibitemShut {NoStop}%
\bibitem [{\citenamefont {Moriya}\ and\ \citenamefont
  {Takimoto}(1995)}]{T.Moriya_JPSJ_1995}%
  \BibitemOpen
  \bibfield  {author} {\bibinfo {author} {\bibfnamefont {T.}~\bibnamefont
  {Moriya}}\ and\ \bibinfo {author} {\bibfnamefont {T.}~\bibnamefont
  {Takimoto}},\ }\href@noop {} {\bibfield  {journal} {\bibinfo  {journal} {J.
  Phys. Soc. Jpn.}\ }\textbf {\bibinfo {volume} {64}},\ \bibinfo {pages} {960}
  (\bibinfo {year} {1995})}\BibitemShut {NoStop}%
\end{thebibliography}

%

\end{document}